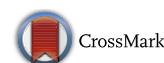

# Stability enhancement of ITO-free non-inverted PTB7:PC$_{71}$BM solar cell using two-step post-treated PEDOT:PSS

Mehrdad Kankanan[1] · Abdolnabi Kosarian[1] · Ebrahim Farshidi[1]

## Abstract

The conductivity and stability of specially treated PEDOT:PSS thin films are investigated. Based on the proposed treatment method, ITO-free PTB7:PC$_{71}$BM organic solar cells are fabricated and the electrical properties of the cells are analyzed. It is shown that by applying a two-step post-treatment method using methanol and ethylene glycol, the conductivity of the PEDOT:PSS thin film increases from 0.9 to 1448 S/cm, and at the same time, a significant improvement of the stability of the layer over time is achieved. It is shown that after 30 days of aging under high humidity condition, the conductivity remains above 70% of its initial value, which is a remarkable result compared with the results reported in the literature. In this paper, important factors affecting the conductivity and stability of the treated layer are studied in detail. In addition, the effect of immersion time in methanol on the conductivity of the layer is also investigated and it is found that dipping times less than 3 min have no appreciable effects on the improvement of the conductivity. An ITO-free non-inverted PTB7:PC$_{71}$BM solar cell is also fabricated using the proposed post-treated PEDOT:PSS thin films as the transparent anode. The power conversion efficiency of the resulting cell is 5.8%. The stability of the fabricated ITO-free cells is considerably better than the stability of the non-treated ITO-free cells or the cells made using ITO as anode.

## 1 Introduction

A single junction solar cell generally consists of an active region, a reflective back contact, and a transparent conductive front contact [1–3]. The transparent front electrode should not only have high conductivity, but also high transparency in the range of absorbing light spectrum, which limits the choice of appropriate materials. Most materials used as the transparent contact are ITO [4, 5], AZO [6, 7], FTO [8, 9] and ZnO [10, 11] among them ITO is the most used material owing to its high conductivity and transparency [3, 12]. However, available indium resources on the earth are limited [13, 14]. Moreover, ITO thin layers are fragile, and flexible solar cells are not possible to fabricate if ITO layers are to be used in their structures [12]. Its poor compatibility with organic materials and

✉ Abdolnabi Kosarian
a.kosarian@scu.ac.ir

1 Department of Electronic and Electrical Engineering, Shahid Chamran University of Ahvaz, Ahvaz, Iran

polymers [12, 15], and transparency degradation at high temperature hydrogen plasma processing conditions are other concerns regarding ITO layers [16]. Therefore, alternative materials such as carbon nanotubes [17–19] and graphene [20, 21] have been under investigation, among them PEDOT:PSS has shown very good properties such as high mechanical flexibility, high optical transparency in the visible range of spectrum, good temperature stability, and high solubility in water, leading to the possibility of thin layers fabrication using solution-based techniques [12, 22, 23]. Nevertheless, the low conductivity of the as-deposited PEDOT:PSS layers and also its conductivity degradation with time are the main drawbacks of this material [12, 24]. A large amount of researches have focused, so far, on the improvement of conductivity of the as-deposited PEDOT:PSS layers, which can be categorized into two main groups: techniques based on pre-deposition modification of aqueous PEDOT:PSS solutions [25–27] and techniques based on post deposition treatments of the deposited thin layers [24, 27–30]. In the first group of techniques secondary dopants such as polar organic solvents [22, 31], anionic surfactants [26, 32], acids and alkalis [33, 34], zwitterions [35], reductive reagents [36, 37], or their combinations are added to the aqueous PEDOT:PSS solution. In the second group of techniques a thin layer of PEDOT:PSS is first deposited on an appropriate substrate and subsequently a treatment using the above mentioned solutions is applied [13, 22–24, 29, 30].

It has been shown that the conductivity improvement of PEDOT:PSS thin layers using post-treatment by high boiling-point polar organic solutions, such as ethylene glycol (EG) and dimethyl sulfoxide (DMSO), leads to similar results as when the solutions are dissolved in the aqueous PEDOT:PSS solution before deposition [38]. But, the improvement of post-treatment conductivity is different, with respect to the pre-treatment method, if other materials such as alcohols, salts, and acids are used [24, 36]. For such materials the post-treatment technique not only leads to better results, but also the removal of extra chemicals from the thin PEDOT:PSS layer is also possible [38]. On the other hand, it is shown that the conductivity of PEDOT:PSS thin layers can be improved to 1362 S/cm using either dipping or dropping methanol post-treatment technique, which shows a remarkable improvement compared with the results obtained using the methanol pre-treatment method [24]. In addition, the methanol treatment is advantageous over the EG treatment from the viewpoint of stability of the conductivity over time [24]. Recently, using a combination of EG and methanol with a ratio of 1:1 could improve the conductivity to 1334 S/cm, which is better than the results obtained using EG or methanol, individually, but no information regarding the stability of conductivity of the layer has been reported [39].

Although most of recent researches regarding improvement of the conductivity and stability of PEDOT:PSS layers have been made using sulfur-based acids [13, 23, 37, 38], but if a PEDOT:PSS layer is to be used as the intermediate layer in a tandem cell, the application of acid-treatment is not possible since acid-treatment can destroy the sub-layers. On the other hand, the amount of improvement in the conductivity depends on the drying temperature of the sample and since the optimum drying temperature in the acid treatment method is usually around 160 °C, the choice of underlying layers encounters limitations [13, 23]. It is also shown that in the acid-based treatment method some acid anions remain in the treated PEDOT:PSS structure [13].

In this paper it is shown that using a simultaneous treatment by methanol and EG, the conductivity of the PEDOT:PSS layer is increased and at the same time its stability with time is also improved.

# 2 Experimental details

## 2.1 Preparation of PEDOT:PSS anode

Although the dropping method has better impacts on the improvement of the conductivity of PEDOT:PSS layers compared with the dipping method [24], but as previously described, it is necessary to dry the samples at an optimum temperature. Therefore, in this research the dipping method is used in the preparation of the PEDOT:PSS layers. The samples are prepared in several steps. At first, a quartz substrate is cleaned in an ultrasonic bath using acetone, methanol and DI water. A thin layer of PEDOT:PSS (PH1000) with high conductivity is deposited using the spin coating technique at 5000 rpm and spin time of 40 s. After deposition, the samples are dried on a hot plate at 130 °C for 20 min. In the next step, the samples are dipped in methanol for 10 min and subsequently in EG for another 10 min. After rinsing the samples in DI water, they are dried on a hot plate at 130 °C. Another set of samples are made and treated with the same method, except for the sequence of treatments, i.e., they are first dipped in EG and then in methanol each for 10 min. For comparison reasons, some samples are also dipped in EG for 20 min, so that the total treatment time for all the samples are the same.

The sheet resistance of the PEDOT:PSS thin films before and after treatment are measured by four point probe technique using gold probes, a Time electronics 2003S D.C Voltage Calibrator and a Time electronics 5075 7.5 Precision Digital Multimeter. Transmission and absorption spectrum of the films are measured using an Analytik Jena SPECTORD S 600 ultraviolet and visible spectrophotometer. The values of transmittances reported here are at the wavelength of 550 nm and include the absorption of the quartz substrate. Fourier transform infrared spectroscopy(FTIR) analyses were obtained using a PerkinElmer Spectrum Two FTIR spectrometer. A DME DS 95 AFM Scanner is used in the ac mode to take the atomic force microscopy (AFM) images of the polymer films and measure their thicknesses. The SEM images are taken using a LEO-1455PV scanning electron microscope.

## 2.2 Fabrication of the solar cells

The ITO-free organic solar cells used in this research are made using Poly[[4,8-bis[(2-ethylhexyl)oxy]benzo[1,2-b:4,5-b']dithiophene-2,6-diyl][3-fluoro-2-[(2-ethylhexyl)carbonyl] thieno[3,4-b] thiophenediyl]] (known as PTB7): [6]-Phenyl-C71-butyric acid methyl ester (known as PC$_{71}$BM). In addition, a number of PTB7:PC$_{71}$BM solar cells using ITO as the anode contact are also fabricated for comparison reasons. The structures of the cells are shown in Fig. 1. The organic photovoltaic (OPV) solar cells are fabricated in several steps described below.

In this research glass laminates used as the cell substrates are first sonicated in an ultrasonic bath using a detergent, acetone, isopropanol, and a solution of 7 wt% of sodium hydroxide (NaOH) in water, each for 9 min. The glass substrates are then rinsed in hot de-ionized water, dried with N2 gas, and heated on a hot plate at 130 °C for 15 min (exposed to air). For the ITO-free OPVs, the PEDOT:PSS aqueous solution (Clevios PH1000) is first filtered through a 0.45 μm PVDF syringe filter. The solution is then coated on the substrate using the static dispense spin coating technique at 3000 rpm for 45 s. The resulting PEDOT:PSS coated glasses are then dried on a hot plate at 130 °C for 20 min in free air. This process is repeated until a thin PEDOT:PSS layer of 76 nm thickness is obtained. In the next step the MeOH–EG post-treatment procedure described in the previous section is applied to the samples. All the samples are stored in a glove box over night and reheated at 130 °C for 3 h before depositing the active layer. The resistivity of the ITO-coated glass substrates used for the OPVs with ITO anode cellsis less than 15 Ω/□ and the transparency is higher than

85% at the wavelength of 550 nm. The ITO coated glasses are first chemically etched with hydrochloric acid using a nylon tape as the mask, and then cleaned as previously explained. After cleaning, a less conductive PEDOT:PSS (Clevios AI 4083), served as a buffer layer, is spin coated on the ITO surface at 5500 rpm for 40 s. Subsequently, the active layer is deposited and baked at 130 °C for 20 min. The active layer solution containing PTB7:PC$_{71}$BM (1:1.5) is prepared in CB with 3 wt% 1,8-diiodooctane (total concentration 15-mg/ml). To completely dissolve the residual polymer, the solution is sonicated in an ultrasonic bath for 2 h at 50 °C, and kept for 1 h in an inert ambient in the glove box. The prepared solution is spin coated on top of the samples at 1600 rpm for 40 s, and immediately placed into a tightly fitting petri dish for about 30 min at room temperature, where the solution is completely dried. Figure 2 shows the normalized (to the highest value) absorption spectra of the PTB7, PC$_{71}$BM and PTB7:PC$_{71}$BM (1:1.5) blends at room temperature, indicating that the desired cell absorbs the wavelengths below 800 nm.

In the final preparation step, using the thermal evaporation technique in a vacuum of $10^{-6}$ Torr, a 1.5 nm-thick LiF layer and, subsequently, a 200 nm-thick Al layer are deposited on the samples. At the end, the cells are vacuum-annealed ($10^{-3}$ Torr,) at 130 °C for 15 min.

For measuring the current density–voltage curves of the cells a Nanosat sun simulator model IIIS-200, calibrated by a standard silicon solar cell, as the light source and a Keithley 2400 source meter are used. For external quantum efficiency (EQE) measurement, the light passes through a monochromator (Oriel Instruments, Model74100) and subsequently is focused on the organic cells under test.

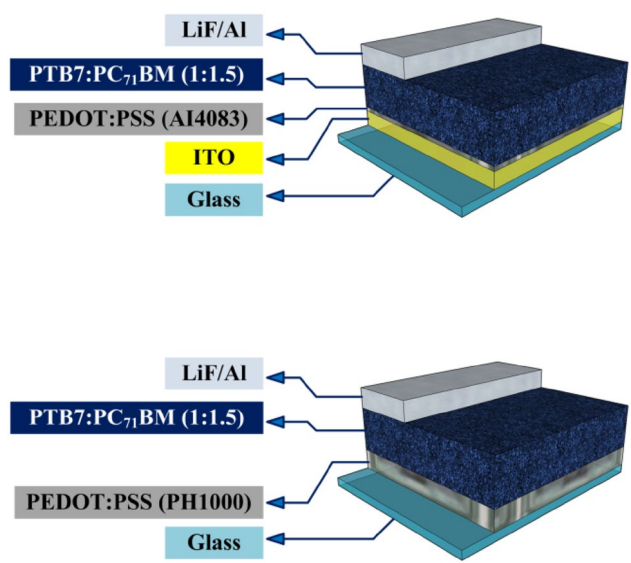

**Fig. 1** Structures of the OPVs with (up) and without (down) ITO anode

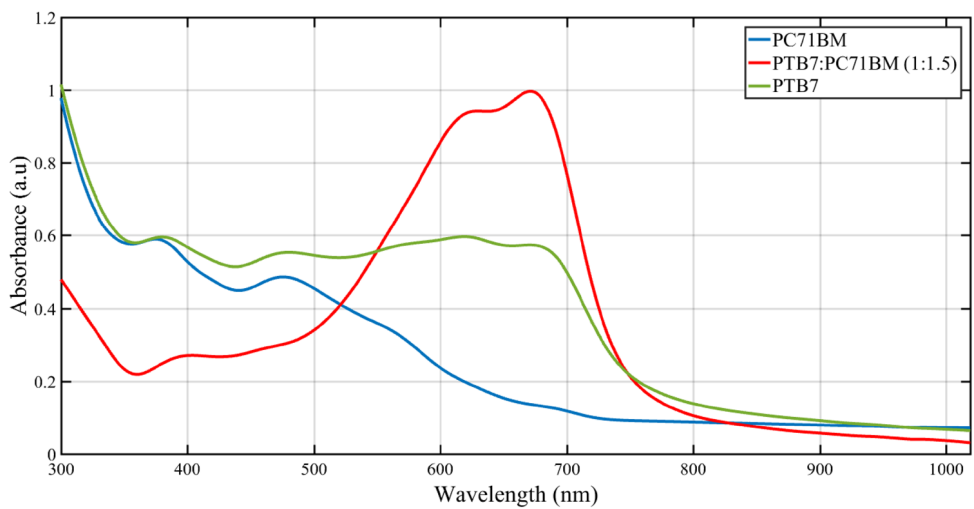

**Fig. 2** Absorbance spectra of PTB7, PC$_{71}$BM and PTB7:PC$_{71}$BM blend (1:1.5) thin films

## 3 Results and discussion

In this paper a two-step treatment is proposed to improve the conductivity and stability of the PEDOT:PSS thin layers. The method consists of dipping in ethylene glycol and methanol in different orders. In order to investigate the effect of dip time on the conductivity, a number of PEDOT:PSS coated substrates are dip-treated in methanol for different times and the conductivity of the layers are measured. The results are shown in Fig. 3, indicating that the dip time could affect the conductivity of the layer, which is in contrast with the results reported in a recent article [29], claiming that the dip time has limited effect on the conductivity. Figure 3 shows that although the treatment time below 1 min has no clear effect on the conductivity, the treatment between 1 and 3 min shows a relatively sharp increase in the conductivity. Above 3 min the conductivity remains nearly constant within an error of ~ 10%. For comparison, some samples are also dip-treated for 5 min in methanol and 5 min in EG. No obvious different was observed between the results.

The conductivity of the samples in response of the orders of solution treatments is shown in Fig. 4. As shown, the post-treatments using EG–MeOH, MeOH–EG and EG–EG sequences, result in the conductivity of the PEDOT:PSS thin layers to improve to about $1370 \pm 110$, $1448 \pm 101$ and $952 \pm 92$ S/cm, respectively. It is commonly believed that the reason for the conductivity improvement is due to two main phenomena happened during the treatments. PEDOT is a highly hydrophobic polymer that is not soluble in the pure water. By adding PSS to the PEDOT as the counter anion, a chemical polymerization phenomenon occurs and the resulting product can be dissolved in water [12]. This is due to the attachment of short PEDOT chains to the long PSS chains by columbic forces.

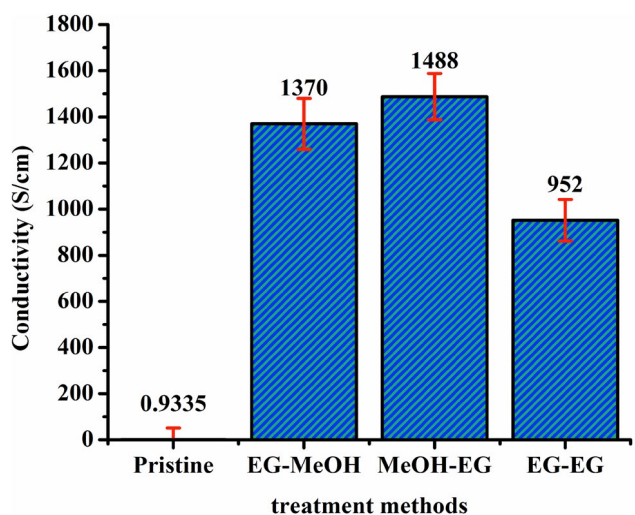

**Fig. 4** Conductivity of PEDOT:PSS thin films treated with different sequence of solvents by dipping method

On the other hand PSS is a hydrophilic polymer; therefore, a core–shell structure is formed in which the core PEDOT chains are surrounded by the shell PSS chains. The columbic repulsive forces among the core–shells make them apart from each other and the portion of the PSS chains that lies between the two shells is converted into a linear conformation. The resulting structure of the PEDOT:PSS dispersed in water is, therefore, in the form of a necklace [12, 23], as shown in Fig. 5.

Since the PSS shells are electrically insulating, and the charge transport is governed across the PEDOT chains, the necklace conformation and coil-shaped PEDOT chains form potential barriers impeding the motion of the free charges and resulting in the conductivity of the initial PEDOT:PSS thin film to be very low, namely less than 1 S/cm [24, 28]. In order to overcome the potential barriers and, consequently, increase the conductivity of the as-deposit thin film, the PSS chains must be removed from the film and the coil-shaped PEDOT must be converted to a linear configuration. This is what essentially happened on the PEDOT:PSS thin films when organic polar solvents and alcohols post-treatments are applied [24, 27].

When an as-prepared PEDOT:PSS thin film is immersed into a highly hydrophilic alcohol, such as methanol, or an organic polar solvent with high boiling point, such as ethylene glycol, an screening effect due to the interactions with hydrophilic PSS chains occurs, resulting in a phase separation between the PEDOT and the PSS chains. Therefore, the PSS chains will be dissolved by the solvents and the orientation of the PEDOT polymer chains rearranges from the coiled to linear or extended-coil structure. This allows more inter-chain interactions and eventually decreases the potential barrier between the conducting polymers, resulting in an increase in the conduction of the PEDOT:PSS thin film [24, 30], In order to support this justification, the UV–visible absorption/transition spectrum in the range of 180–300 nm is investigated. The result is shown in Fig. 6. Since the two absorption bands in the ultraviolet range are related to the benzene rings in PSS, the decrease in the absorbance peaks over the range of the spectrum confirms the removal of PSS chains from the initial thin film [12, 24]. The same conclusion is also made from the transmission spectrum.

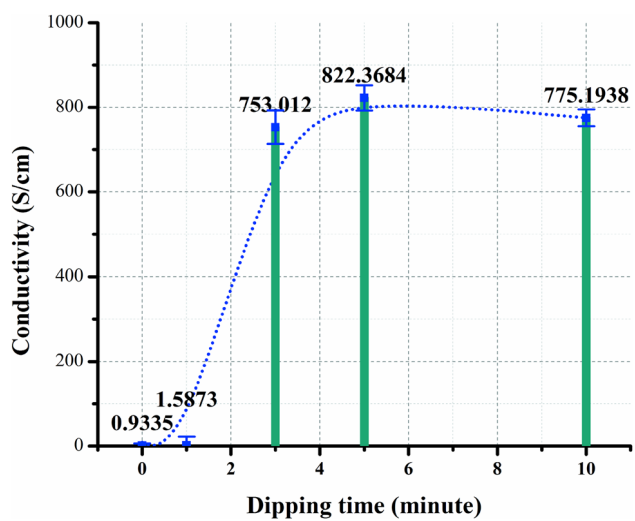

**Fig. 3** Effect of methanol immersing time on conductivity enhancement

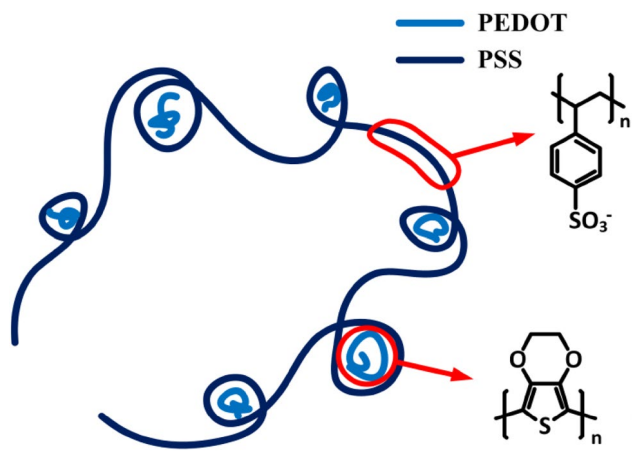

**Fig. 5** PEDOT:PSS necklace-like structure in aqueous solution

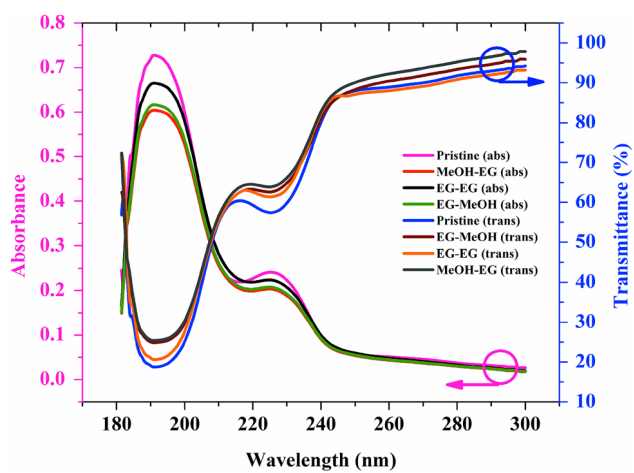

**Fig. 6** UV–visible absorbance/transmittance spectra of PEDOT:PSS film before and after treatment

FTIR analysis achieved on the samples also supports these results. Two types of samples are used for this experiment. The first type contains the residuals of the solutions remained after the thin film post-treatment. FTIR measurements are conducted by dropping these solutions separately on the KBr cells with a ratio of 1:20 in weight. The resulting FTIR spectrum is presented in Fig. 7. The absorbance peaks at 998 and 1035 cm$^{-1}$ correspond to S=O symmetric, and the absorbance peaks at 1125 and 1169 cm$^{-1}$ correspond to S=O asymmetric stretching vibration in the PSS [24, 40]. The second types of the samples are the thin films formed on glass substrates and treated with different methods. The results of FTIR spectroscopy of these samples in the range of 500–1500 cm$^{-1}$ are shown in Fig. 8, confirming the existence of important absorbance peaks of –SO$_3^-$ functional group in the PSS.

Comparing the results in Figs. 7 and 8 it can be concluded that:

- There are some trace of PSS particles in the solutions indicating the reduction in the amount of PSS in the PEDOT:PSS thin film.
- The intensity of the peaks at 998, 1035, 1125 and 1169 cm$^{-1}$, corresponding to the –SO$_3^-$ group in the PSS, obtained from the MeOH–EG and EG–MeOH methods is higher compared with peaks obtained from the EG–EG method, which indicates the removal of more PSS from the thin film using these methods and subsequently the improvement of the conductivity of the films.
- The reduction of the intensity of the peaks at 1006 and 1035 cm$^{-1}$ confirms the reduction of the amount of PSS in the thin film.
- Comparing the magnitude of the peaks in Fig. 8 reveals that the MeOH–EG, EG–MeOH and EG–EG methods are, respectively, more effective in the PSS content removal, therefore, the conductivity obtained using the MeOH–EG method should be more than the other post-treatments used in this work. This result is in agreement with the result illustrated in Fig. 4.

AFM images presented in Fig. 9 are used to investigate possible differences in the morphology of the films before and after treatments and also the correlation between the morphology and the conductivity. As observed in Fig. 9a and e, the pristine PEDOT:PSS thin films, having low conductivity (less than 1 S/cm), consists of irregular PEDOT (bright regions) and PSS (dark regions) chains with no distinct separation between their phases. As the post-treatment solvent is applied to the layer, the separations between the PEDOT and the PSS chains increase, the PEDOT chains grow in size, and the interconnections between conductive PEDOT polymers become more efficient. The average ($S_a$), mean value ($S_m$) and root mean square ($S_{rms}$) roughness, calculated from the AFM data, are summarized in Table 1. It indicates that all layers have acceptable low surface roughnesses from 1.97 nm for the pristine film to 2.75 nm for the MeOH–EG post-treated film. Of course, the post-treatment process increases the roughness of the films, because of the larger size of the PEDOT grains.

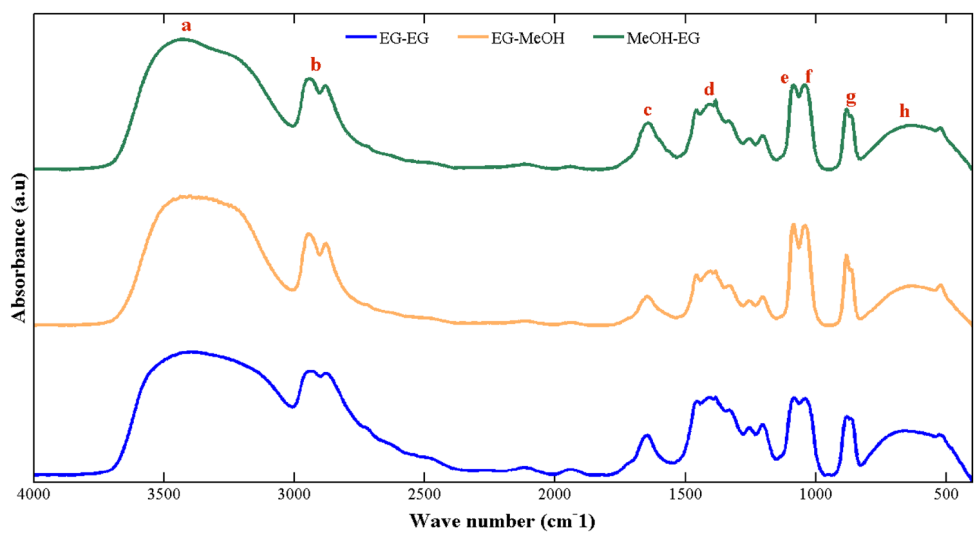

**Fig. 7** FTIR spectra of the solutions remaining after the PEDOT:PSS thin film post-treatment. *a* 3000–3700 cm⁻¹: water O–H groups absorbed in the sample (sulfonic acid groups are strongly hydrophilic), *b* 2934,2880 cm⁻¹: alkanes C–H stretching vibration, *c* 1643 cm⁻¹: O–H bending vibrations, *d* 1406,1458 cm⁻¹: aromatic C=C stretching vibrations, *e* 1125,1169 cm⁻¹: S=O asymmetric stretching vibrations, *f* 998,1035 cm⁻¹: S=O symmetric stretching vibrations, *g*: 658, 833 cm⁻¹: =C–H out of plane deformation vibrations, *h* 527 cm⁻¹: ring in-plane deformation variations [24, 40]

**Fig. 8** FTIR spectra of PEDOT:PSS post-treated thin films with different methods in the range of 500–1500 cm⁻¹. Oscillatory vibrations at x = 1006 and y = 1035 are due to –SO₃⁻ group in PSS

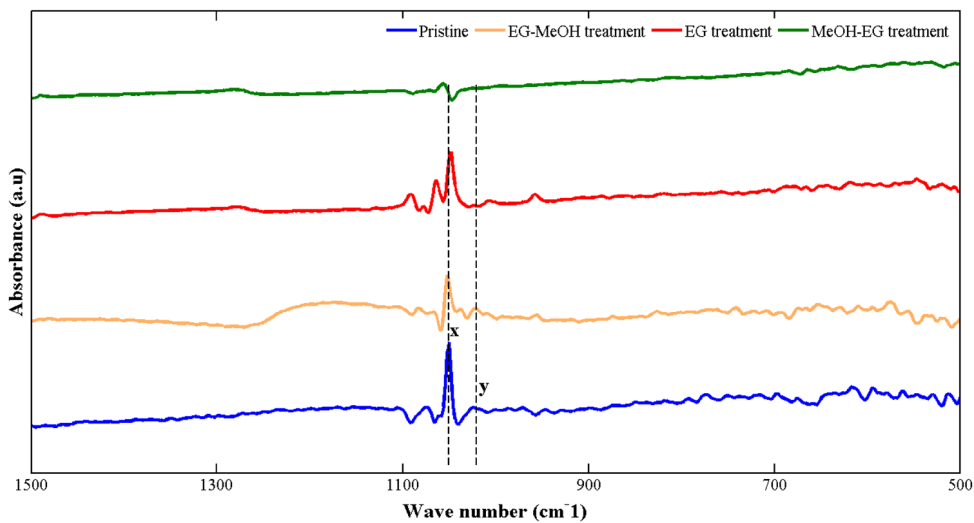

Another technique used for investigating the layers is scanning electron microscopy. The scanning electron microscope (SEM) images obtained from the PEDOT:PSS layers prepared by spin-coating method on pre-cleaned (using acetone, isopropyl alcohol, and DI water) aluminum foils at 3000 rpm for 45 s are provided in Fig. 10. As shown in Fig. 10a, the surface of the pristine PEDOT:PSS layer is quite smooth and uniform in an area of 2×2 μm, but after the post-treatments, sub-micrometer particles appear in the PEDOT films (Fig. 10b–d). These particles can be rinsed from the surface using water or methanol. It seems that the PSSH chains are expelled from the thin layer by the screening effect during the process of drying with methanol or ethylene glycol.

The current density–voltage (J–V) curvesunder the standard AM1.5G solar simulator illumination (100 mW/cm²) and external quantum efficiency (EQE) of the fabricated solar cells are presented in Fig. 11a and b. Furthermore, the average short circuit current density (J$_{SC}$), open circuit voltage (V$_{OC}$), fill factor (FF), power conversion efficiency (η), series resistance (R$_S$) and parallel resistance (R$_{Sh}$) of the OPVs are presented in Table 2. According to the results, the solar cells with PEDOT:PSS post-treated anode have similar performance as the solar cells with ITO transparent electrode.

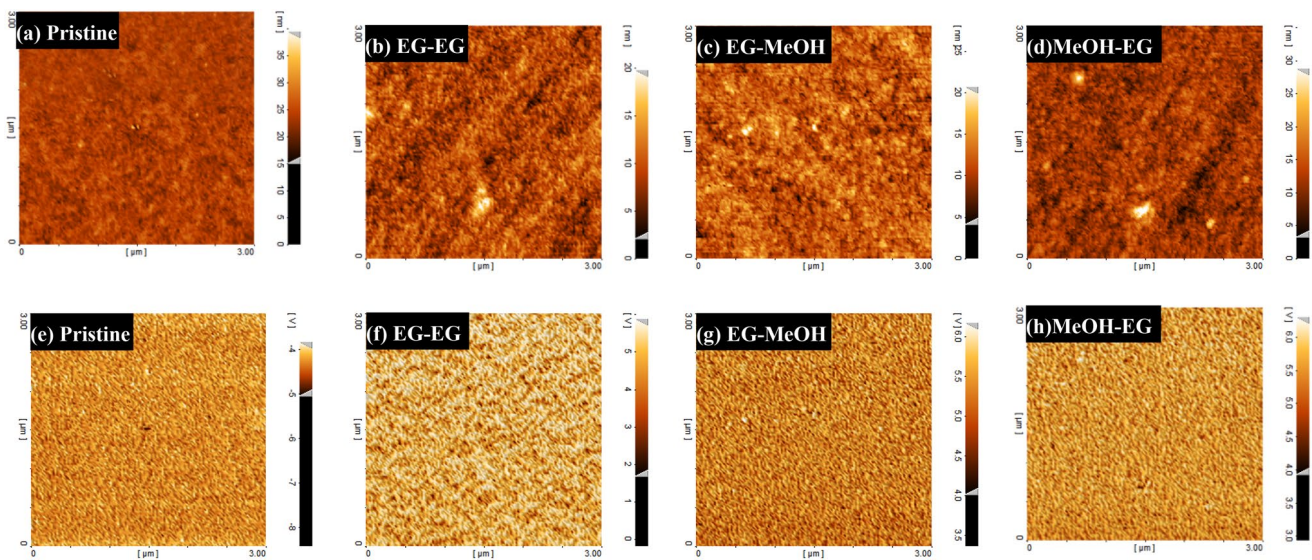

**Fig. 9** AFM images in ac mode showing the surface morphology (**a, b, c, d**) and phase images (**e, f, g, h**) of pristine, EG–EG, EG–MeOH and MeOH–EG post treated PEDOT:PSS thin films. Scanning area: 3×3 μm

**Table 1** Average ($S_a$), mean ($S_m$), and root-mean-square ($S_{rms}$) roughness of PEDOT:PSS layers prepared by different methods

|  | Pristine | EG–EG | EG–MeOH | MeOH–EG |
|---|---|---|---|---|
| $S_a$ (nm) | 1.57 | 2.04 | 1.86 | 2.19 |
| $S_m$ (pm) | 5.15 | 4.97 | 5.48 | 5.38 |
| $S_{rms}$(nm) | 1.97 | 2.54 | 2.34 | 2.75 |

The solar cells with ITO anode have a short circuit current density of 13.8 mA/cm², an open circuit voltage of 0.68 V, fill factor of 67% and an average power conversion efficiency (η) of 6.3% which are comparable to the reported values in the literature for PTB7:PCBM non-inverted OPVs [41–44]. Therefore, it is used as a good reference for evaluating the ITO-free OPVs. The OPVs with post-treated PEDOT:PSS transparent anode have less power conversion efficiency (5.8%) and fill factor (63.1%), due to higher series resistance, and have lower short circuit current density (13.52 mA/cm²), due to their higher shunt resistance. In addition, according to the optical properties of the PEDOT:PSS thin film (Supplementary Information Figs. S I, II), it is clear that by increasing the thickness of the layer and also decreasing the density of the PSS chains in the thin film (in the case of the post-treated PEDOT:PSS transparent anode), the absorbance of the PEDOT:PSS electrode increases in the visible and IR range of the solar spectrum, and accordingly the transmittance is reduced, which reduces the short-circuit current of the cells. This is also shown in the EQE curves for the two OPVs with post-treated PEDOT:PSS and with the ITO transparent anode.On the other hand the open circuit voltages in either cases are approximately equal, with a slightly higher $V_{OC}$ (0.71 V) for some of the ITO-free cells.The PEDOT:PSS treatment process reduces the defects in the layer and its interface with the active area [39], which allows better charge transfer across the interfaces and increases FF and $J_{SC}$. Therefore, FF, $J_{SC}$ and consequently the power conversion efficiency (η) of the solar cell with un-treated PEDOT:PSS electrode are less than those of the other cells. Table 2 also shows the amounts of $R_S$ and $R_{Sh}$ of the cells. The series resistance of a solar cell consists of the contacts resistance and the active bulk layer resistance. The difference in $R_S$ of the three cells may results from the difference in the conductivity of the transparent contacts, the interface between transparent anode and active area and the charge carrier mobility in the transparent contact thin film [45].

An important issue in the performance of a solar cell is its stability when exposed to the free air. Because of the effects of oxygen, humidity and ultraviolet radiation on the stability of the polymer layers, organic solar cells usually have short life span in air [42, 44, 46]. One of the main reasons for degradation of the performance of the polymer solar cells made from PEDOT:PSS layers, is the change in the resistance of the underlying PEDOT:PSS films. In this work the stability of the PEDOT:PSS thin films before and after treatment with different methods and their overall effects on the OPV performance have been investigated during a large period of time. Figure 12 shows the changes in the conductivity of different PEDOT:PSS layers, held in the free air at 37 °C and humidity more than 70% for 30 days after fabrication. It is observed that after 30 days, the conductivity of the PEDOT:PSS thin films with no treatment drops to 17% of its initial value. This results in a substantial reduction in the solar cell efficiency. On the other hand,

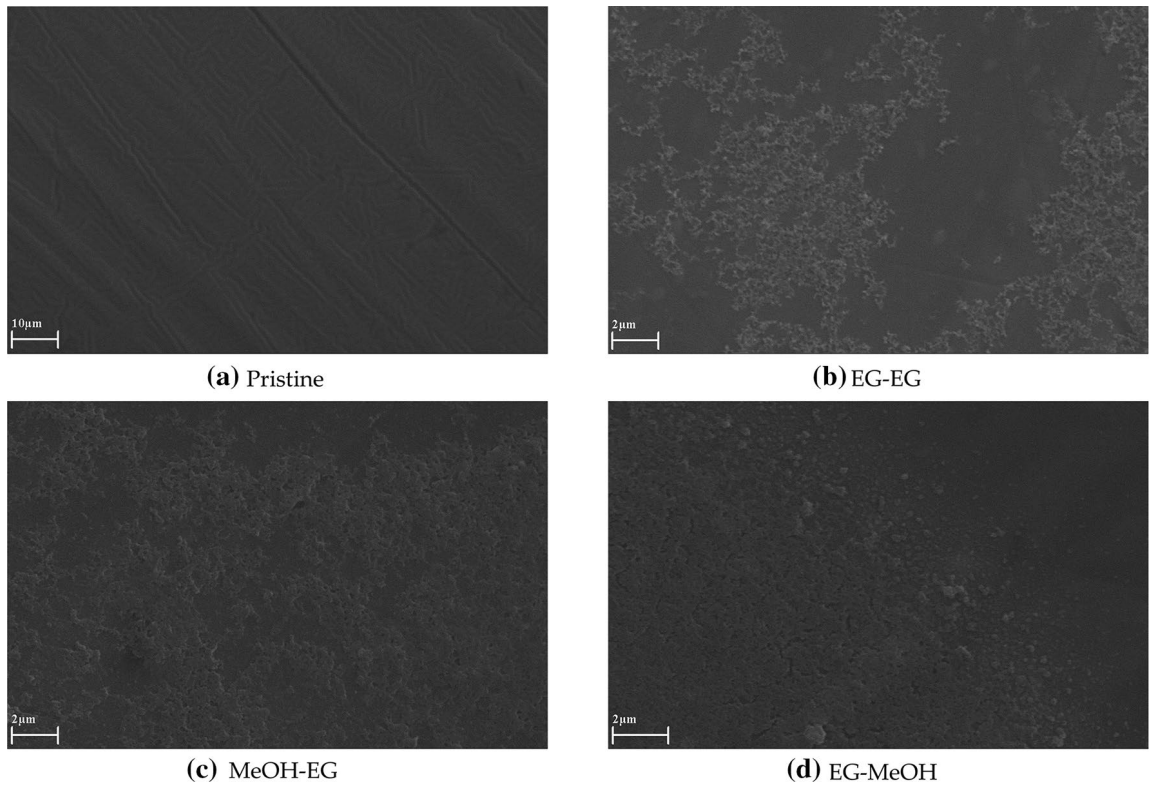

**(a)** Pristine  **(b)** EG-EG

**(c)** MeOH-EG  **(d)** EG-MeOH

**Fig. 10** SEM images of pristine, EG–EG, EG–MeOH and MeOH–EG post treated PEDOT:PSS films

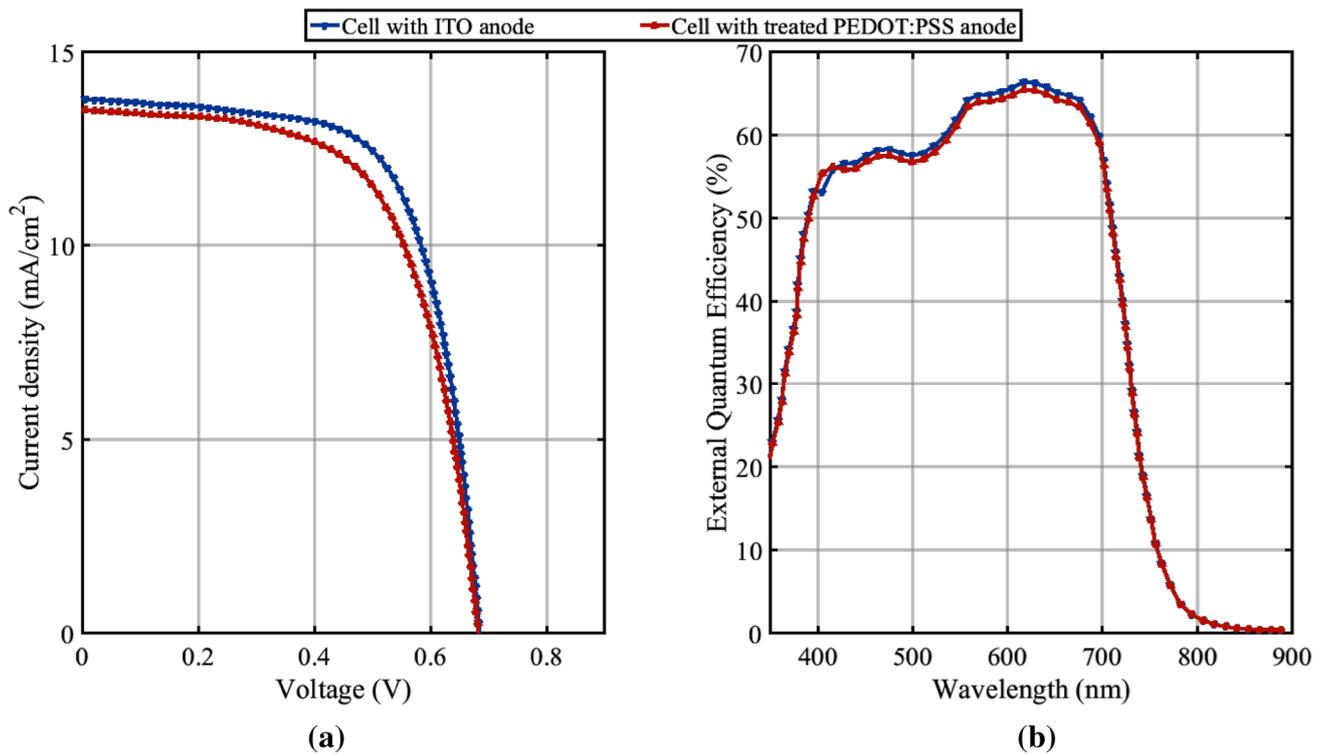

**(a)**  **(b)**

**Fig. 11** **a** Current density–voltage curves and **b** external quantum efficiency spectra of the OPVs with (blue) and without (purple) ITO anode. (Color figure online)

EG–EG, EG–MeOH and MeOH–EG treated films are more stable where the conductivity drops to 60, 54, and 73% of the initial values, respectively. On the other hand, according to Fig. 13, the transparencies of the layers are above 88% over the range of visible spectrum (higher than 92% at the wavelength of 550 nm). It is observed that the two step

MeOH–EG treatment on the PEDOT:PSS layer gives the best results from the viewpoint of the conductivity of the anode contact.

In the next step, three sets of cells are fabricated each set consisted of 8 cells of equal effective active layer area (0.15 cm²). The structure of the first set of the cells is

**Table 2** Parameters of the OPV cells with ITO or PEDOT:PSS post-treated anode

| Transparent Anode | $J_{SC}$ (mA/cm²) | $V_{OC}$ (V) | FF (%) | $\eta$ (%) | $R_S$ ($\Omega$ cm²) | $R_{Sh}$ ($\Omega$ cm²) |
|---|---|---|---|---|---|---|
| ITO | 13.8 | 0.68 | 67 | 6.3 | 6.7 | 1268 |
| Post-treated PEDOT:PSS | 13.52 | 0.68 | 63 | 5.8 | 9.1 | 1315 |
| Un-treatedPEDOT:PSS | 0.27 | 0.68 | 31.5 | 0.05 | 34 | 5435 |

**Fig. 12** The stability of the pristine and treated PEDOT:PSS thin films with different methods

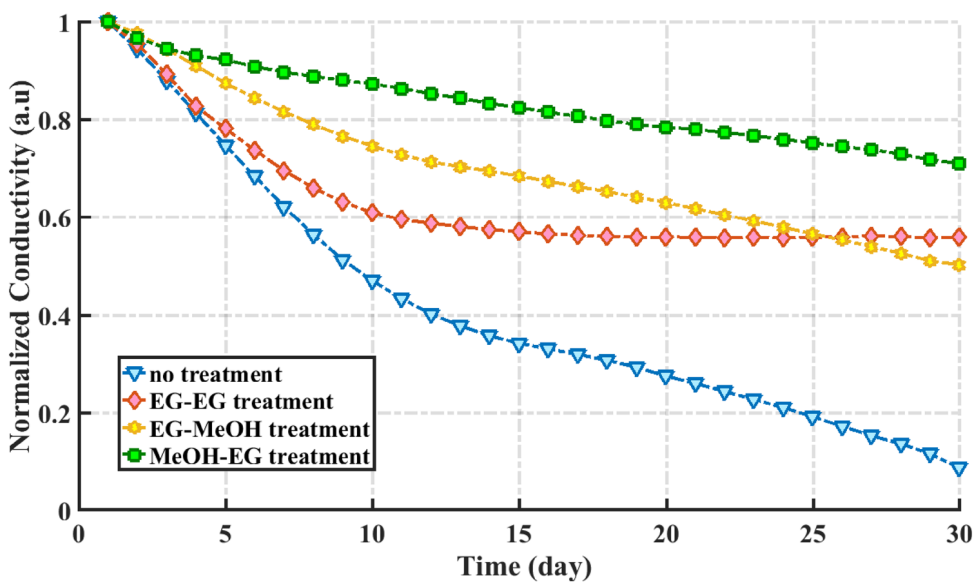

**Fig. 13** Transmittance of 17 nm-thick AI 4083 and 30 nm-thick PH1000 PEDOT:PSS treated layer with different methods spin coated on glass

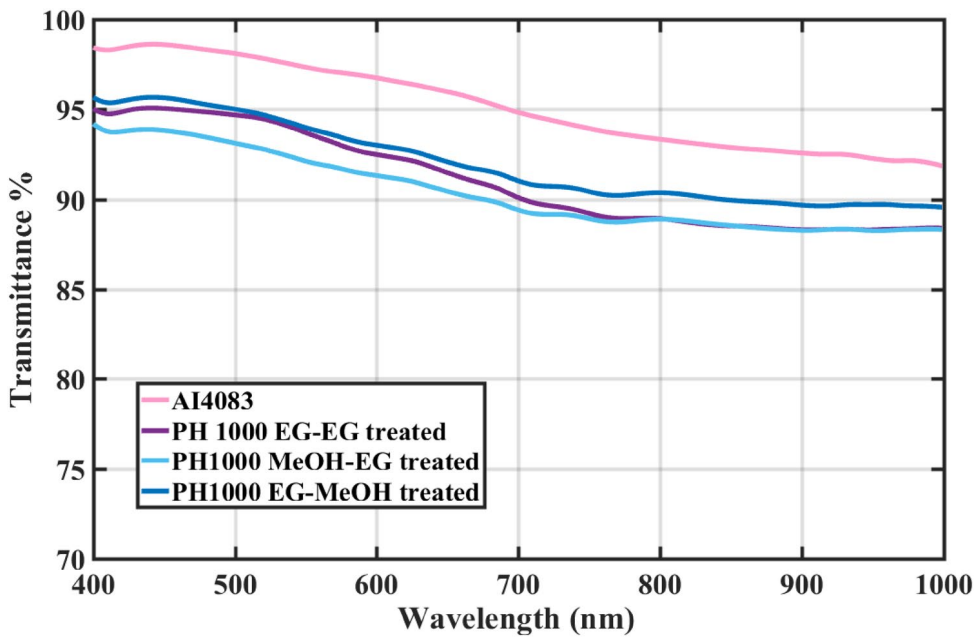

untreated-PEDOT:PSS (PH1000)/PTB7:PC$_{71}$BM/LiF/Al (cell-1), the second set is MeOH–EG–treated PEDOT:PSS (PH1000)/PTB7:PC$_{71}$BM /LiF/Al (cell-2), and the third set is ITO/PEDOT:PSS (AI 4083)/PTB7:PC$_{71}$BM /LiF/Al (cell-3). After fabrication, the cells are placed in the free air ambient at temperature of 27 °C and humidity of 37% in the dark, and are exposed to the light only during the measurements achieved during a time span of 1300 h. To minimize the effect of light soaking on the performance of the cells, the time of light exposure and measurements is forced to be as short as possible. The results shown in Fig. 14 are the normalized values obtained from the measurements on eight cells of each set, having in mind that the initial values of each set are previously reported in Table 2.

The reduction in the open circuit voltage (V$_{OC}$) is 6.56, 3.81, and 11.51% for the cell-1, cell-2 and cell-3 sets, respectively. Regarding the efficiency and J$_{SC}$, it is observed that the cell-1 sets show no changes in the first 40 min, after which a nearly exponential drop is observed. For the cell-2 set, the efficiency and J$_{SC}$ are stable in the first 110 min and then the efficiency drop is seen at a lower rate than cells-2 and cell-3 sets. The decrease in efficiency for cells-3 sets is visible from the beginning, so that after 255 min the efficiency is dropped to less than 60% of its initial value. As a result it is concluded from the curves that the set 2 of the samples are the most stable cells, where their efficiency remains above 50% of its initial value even after 1035 min of their fabrication. This is a remarkable result since most of the reports regarding the organic cells show a large reduction in the efficiency after a few minutes from their fabrication [46, 47].

It should be noted that the drop in the efficiency of the organic solar cells has various and complex reasons, but what is most concerned about here is the change in the properties of the PEDOT:PSS layer. As hygroscopic PSS$^-$ group absorb humidity from the air, the conductivity of the PEDOT:PSS layer decreases. Moreover, the anode/buffer layer will exhibit an acidic characteristic [48] and diffuse to the other layers [49] which will destroy the active layer (and the underlying ITO anode). Therefore, the existence of more amount of PSS in the PEDOT:PSS layer results in faster degradation of the structure and consequently faster reduction of the cell's efficiency. Oxidation of the Al cathode is another reason for efficiency drop in OPV cells having no encapsulation. In addition, in the cells with ITO anode the reduction rate is faster than the two other cells without ITO

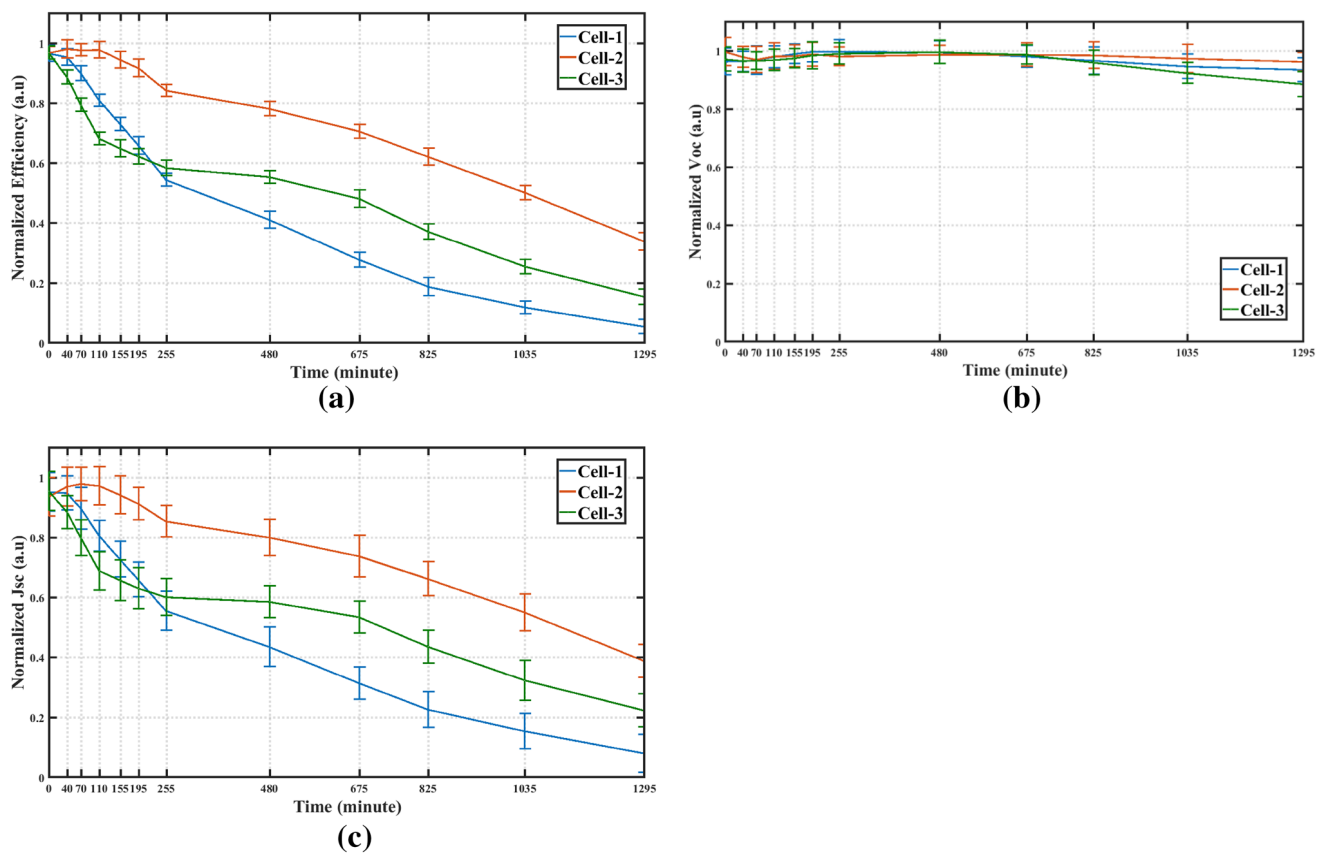

**Fig. 14** Normalized **a** power conversion efficiency, **b** open circuit voltage and **c** short circuit current variations of three sets of cells during 1300 h

anode (in the first 200 min), as explained in [50] it is because of the migration of indium into the PEDOT:PSS buffer layer. But as previously mentioned, the change in the properties of the PEDOT:PSS layer is important here.

## 4 Conclusions

A two-step post-treatment method is proposed to improve the conductivity of as-deposit PEDOT:PSS thin films, consisting of a submersion step in methanol and an immersion step in ethylene glycol in various sequences. It is observed that if the immersion time in methanol is less than 3 min, there will be no significant enhancement in the conductivity of the PEDOT:PSS thin film. On the other hand, if the PEDOT:PSS thin film is first immersed in methanol and subsequently in ethylene glycol, not only the conductivity will increase from 0.93 S/cm to 1448 S/cm, but also its stability will improve compared with the stability of the film treated in just ethylene glycol or methanol. It is shown that keeping the layer in the free atmosphere for a long period of 30 days, the conductivity of the film treated by the proposed method encounters only some 27% reduction with respect to its initial value which is less than the 40% loss in the conductivity of the PEDOT:PSS thin film treated with ethylene glycol. Based on these results and owing to its high transparency, the treated PEDOT:PSS thin film is a very promising alternative for the ITO anode used in organic PTB7:$PC_{71}$BM solar cells. As observed, the proposed ITO-free non-inverted PTB7:$PC_{71}$BM solar cell has an efficiency close to the efficiency of the organic solar cell with ITO anode. Moreover, using the proposed post-treated PEDOT:PSS thin film as the OPV transparent electrode, the stability of the organic solar cell will also improve. It is shown that after more than 1035 h from the cell fabrication, the efficiency of the cell, kept in the environmental condition, remains above 50% of its initial value. This noticeable improvement of the stability of the PTB7:$PC_{71}$BM solar cells over such a large period of time can open new windows to the organic cell technology.